# Optimization of brain and life performance:
# Striving for playing at the top for the long run

Didier Sornette

ETH Zurich
Department of Management, Technology and Economics
Kreuzplatz 5, CH-8032 Zurich, Switzerland
email: dsornette@ethz.ch

**The seven recipes: motivation and background**

Everyone may think she knows what is good and what is bad. One more theoretical article will hardly make anyone act differently. My goal here is to share simple recipes that are easy and often fun to put in practice and that make a big difference in one's life. I am writing this note as an opportunity to share my most important snippets of advices with my students and younger collaborators at ETH Zurich. I hope that professionals and the broader public may also find some use for it, as I have seen already the positive impacts on some of my students.

Why optimize? Why performance? Actually, most of us probably just want to enjoy their life and find meaning to it. I claim that having full body and mind performance, to a degree that most of us do not think possible at a consistent level, is easily achievable and increases greatly the breadth of experiences and the enjoyment of life. As children, we love to run at our top speed, try all kinds of experiments that our mothers would rarely allow, push our limits, and experiment. As teenagers and then young adults, we are eager to use our burgeoning talents and strengths to try new adventures, to sometimes push the limits to extremes. Is it not fantastically more fun to have a stronger and flexible body to try and practice sports of any kind?  Is not music much more enjoyable when mastering more techniques and when exercising a trained ear? Is it not fun to play chess at the Master level? This list extends to all human activities. It should be obvious that the dimension and breadth of enjoyment is amplified, the higher the powers of the body and the mind. But we often accept our limits and settle for what we have, especially when aging whose deteriorating effects are wrongly considered inevitable.

Modern medicine with the revolution of biotechnology, the achievement of the Human Genome Project and the booming disciplines of genomics, proteomics and other "ics" put a lot of emphasis on genes as the cause for diseases as well as psychological quirks, as opposed to environmental and acquired traits. This innate versus acquired debate is a minefield of controversies that reflects the great complexity of biology, evolution and ecological interactions. My interpretation of the enormous existing scientific literature is that the emphasis on genes expresses something akin to the bias of "searching the lost key under the Lamppost", i.e., following a path that is amenable to technical approaches with a framework of linear causal thinking. However, as a specialist of complex systems, I hypothesize that most of the diseases and problems, if rooted in a genetic landscape (this is almost a pleonasm since our genetic code controls the basic instructions of our biology), are actually due to or facilitated by environmental processes that can be predicted and/or controlled. In their careful review of the literature on cancers, Campbell and Campbell [2006] as well as Servan-Schreiber (2009) find that genes constitute only a few percent of the factors that promote cancers. If genes account for a minor part, what we do, drink, eat, perform and think constitutes the dominating factor.  My approach is based on the hypothesis that we can control a large part of our life, of our health, of our performance and success.

I refuse to settle for what our culture considers inevitable, i.e., irregular and/or sub-optimal performance and decreasing abilities with age. I claim that it is possible to enjoy every day to the fullest, to play and work with close to 100% of our peak performance, to be continuously "in the moment" and this for many decades. We should not settle for less.

This essay offers a few recipes to achieve sustained brainpower and body health, strength and resistance. The recipes are so simple as being obvious and almost trivial. However, they are regularly rediscovered and re-forgotten, probably because there are no economic benefits in the short term to the dominant agro-business and pharmaceutical establishment (some visionary companies recognize however the win-win opportunities with investing to develop new products for healthy life styles). And motivations and incentives in our societies are often misguided with emphasis on short-term gains at the expense of sustainability and social responsibility. I claim that the principles outlined below are so simple and straightforward that they can be easily implemented with a minimum of will, organization and time cost. I claim that the results that derive from using these simple recipes are life changing, both for better performance, quality of life, health, happiness and longevity. They also call for a reassessment of what is often taken for granted in our modern life at a broad and deep level.

I am not a professor of medicine, neither of health, nor of nutrition or sport. I am a professor on the chair of Entrepreneurial Risks in the department of Management, Technology and Economics at ETH Zurich, a professor of Finance at the Swiss Finance Institute, the director of the Financial Crisis Observatory ([www.er.ethz.ch/fco](www.er.ethz.ch/fco)), a founding member of the Risk Center at ETH Zurich (June 2011) ([www.riskcenter.ethz.ch](www.riskcenter.ethz.ch)), a professor of Physics associated with the Department of Physics (D-PHYS) at ETH Zurich and a professor of Geophysics associated with the Department of Earth Sciences (D-ERWD) at ETH Zurich. You thus see that I strive to develop interdisciplinary understanding of and solutions for complex problems. The list of my positions also suggests that I spend long intensive hours at the office and need to work probably harder than many mono-disciplinary persons. I have thus always been motivated to find ways to boost my performance.

The recipes presented shortly have been inspired from collaborative ties with colleagues in various medical and health institutions [Sornette et al., 1994; Sornette et al., 2009; Osorio et al., 2009; 2010]. These collaborations reflect my dream of developing an integrative way of life, science and society, as my research might attest. In addition to the standard mentoring of my juniors, as should be expected from my functions, I consider it to be my duty to tell my students and collaborators to take better care of themselves and of their health, so that they may perform better, that they do not become ill at the flu seasons in autumn and spring, and may work more efficiently without losing intensity in their long study hours. I claim that it should be possible to perform at the top, like an Olympic athlete, but with the difference that the goal is to excel over the long run, over a 50 years horizon, and more! I practice daily these simple rules to achieve top brain and professional performance as well as enjoy fully my everyday life.

As I am writing this piece, I am pushed by my older son Paul-Emmanuel (23), a superb athlete with impressive strength and stamina, to reveal that, at 54, I am still beating him in arm wrestling, that I often practice with him some extreme sports, that I am able to work for 12-14 hours at close to peak brain performance everyday, that recently we have with my son between us crossed the strait on the Mediterranean sea between Nice and Corsica wakeboarding behind a speedboat over almost 200 km, that returning from international travels to the other side of the globe, I can perform intense monoski on the lake of Zurich with an outside temperature of 4 C with almost no jetlag and minimal protection and with the enjoyment resulting from the full play of a well-functioning body. This may be seen as bragging. My son insisted that this paragraph should be added to give a concrete example of what can be done easily, without the ordeals that Olympic athletes impose on themselves, that this is the "normal" and typical level of performance and playfulness we could all achieve as a matter of routine by following the simple easy-to-implement recipes outlined below. And for the busy professionals among us, I stress that the time cost is shockingly small (see also [Ferriss, 2011]).

I first state the seven guiding principles and then develop their rationality, expected consequences and describe briefly how to put them in practice: (1) sleep, (2) love and sex, (3) deep breathing and daily exercises, (4) water and chewing, (5) fruits, unrefined products, food combination, vitamin D and no meat, (6) power foods, (7) play, intrinsic motivation, positive psychology and will. These simple laws are based on an integration of evolutionary thinking, personal experimentation, and evidence from experiments reported in the scientific literature. The following inspiring books provide complementary information and many useful references [Robbins, 1997 in chapter 10 [1]; Campbell and Campbell, 2006; Servan-Schreiber, 2009; de Vany, 2010 [2]; Ferriss, 2011].

Let me restate the seven principles with some explanations before giving the full descriptions in the seven following sections.

1) **Sleep**: Rest with quality sleep for a minimum of 7-8 hours per night;
2) **Love and sex**: Make love as often as possible with your special partner and cultivate the romance and relationship; interrupt your work when needed with one minute of intense focus on the loved one, perhaps using romantic or sexy pictures of him/her to trigger happiness hormones that boosts brain performance and well-being.
3) **Deep breathing and daily exercises**: Start each of your day (no exception) with 5-10 minutes of exercises, including deep breathing-stretching followed by abdominal and finishing with a very short intense workout; perform a few 2-3 minutes of intense workouts and deep breathing at different times of your day in your office or wherever you happen to be in order to oxygen your body and refresh your brain;
4) **Water and chewing**: Drink at least 2 liters of water per day (no canned juice, no coke, no beer, no sugar) outside meals and drink minimally or not at all during meals (a small glass of red wine or cup of hot green tea is fine); "drink your food" and "eat your drinks".
5) **Fruits, unrefined products, food combination, vitamin D and sun exposure and no meat and no dairy**: Eat as much fruits with water as possible on an empty stomach during the day, avoid meat and consume only unrefined products and cereals; avoid bad food combination to avoid conflicts between alkaline versus acid foods.
6) **Power foods**: onion, garlic, lemon, kiwis, almonds, nuts, dry fruits for super-performance in time of intense demand.
7) **Play, intrinsic motivation, positive psychology and will**: rediscover the homo ludens in yourself in things small and large so that work and life become a large playground, cultivate motivation as a self-reinforcing positive feedback virtuous circle.

**1-Sleep**

Sleep seems to be an unproductive activity for those of us who would rather play, enjoy company, work or create. But recent research shows that sleep has many benefits in addition to the consolidation of memory and other brain functions. During sleep, the brain sends signals to trigger cellular repair of damage caused by metabolic processes and goes through cortical reorganization to process sensory inputs [Savage and West, 2007]. During sleep, our body and brain goes to a "maintenance" and repair mode, effectively rejuvenating the body. In addition, sleep has been shown to facilitate the process of insight [Wagner et al., 2004] and many of us have experienced bursts of inspiration upon awakening.

I consider sleep the single most important performance step both for the short and long term maintenance of performance. However, the sad truth is that many persons who strive for performance cheat on sleep. This is well documented to lead to decreased efficiency, unhappiness, anxiety, strain and depression as well as a weakened immune system. Adequate sleep provides energy and enthusiasm to embrace the day. Sleeping long and deep is paradoxically a gain of time because our productivity as well as our senses of well-being and alertness are so much improved. Sleep-deprived people often enter a pro-cyclical feedback process that tends to self-perpetuated lack of sleep: more time is needed to accomplish tasks and the less one sleeps, the more difficult it is to find sleep. After a while, sleep-deprived people more or less adjust to their new homeostasis, developing tolerance to the feelings of sleepiness that makes them unaware of their deteriorating alertness and performance. Accepting their new (suboptimal) normal, they compensate the negative effects by various drugs that include coffee and anti-depressors.

---

[1] Chapter 10 is a remarkable synthesis that complements the present discussion. However, I do not necessarily endorse all the other chapters.

[2] Art de Vany does not emphasize enough the need to drink a lot of water and to eat many water filled fruits with dark colors. The recommendation to avoid carbohydrates should be amended by distinguishing between refined (not recommended, I agree) and unrefined unprocessed carbohydrates (which are recommendable). In my experience and opinion, the emphasis on protein meat is misguided.

It is known that older people tend to sleep less. I hypothesize that sleeping less makes people age faster and aging leads to less sleep, which tends to make aging faster, which… goes on in a pro-cyclical feedback vicious circle. I conjecture that keeping a regular sleep of about 7-8 hours per night as age advances ensures conservation of most faculties over the long-term and delays aging considerably by promoting the repair processes in the body. It helps to have regular times for sleep as the body functions a lot on circadian cycles and needs signals to trigger sleep. In my experience, "early birds" tend to have deeper regenerating sleeps than "night owls": try to sleep before 11pm. When severely sleep deprived, I recommend a 5 minutes meditation-like shutting off of the mind with closed eyes, say by isolating you in the bathroom. This replenishes the energy and transforms the rest of the day.

**2-Love and sex**

Achieving continuous high brain performance levels requires a strong personal motivation and reward system. One very efficient channel to ensure that rewards continuously wash the brain is love.

According to H. Fisher [2004, 2009], love is not an emotion but "a motivation system, it's a drive, it's part of the reward system of the brain. It's a need that compels the lover to seek a specific mating partner." She emphasizes that love is the most potent motivator that evolution has given us. Love, which according to H. Fisher can be classified into (i) lust-love, (ii) attraction-love and (iii) attachment-love, is associated with the secretion of very powerful hormones for the body that also act as neurotransmitters. Reward processes in the nervous system seem to be associated with many functions in the body, including the immune system [Ader et al., 1990]. In particular, attraction-love (also called romantic or passionate love) is associated with high levels of dopamine, which plays an essential role in controlling the brain's reward and pleasure centers as well as the regulation of movement and emotional responses. In turn, attachment-love is driven by the hormones oxytocin and vasopressin, which come with a sense of calm, peace, and stability one feels with a long-term partner. Attachment-love also includes "loved ones", such as family, friends, and pets (see "feline and canine therapy" approaches). As for the first love class (love-lust), in a large scale economic study of the links between income, sexual behavior and reported happiness using data on a sample of 16,000 adult Americans, Blanchflower and Oswald [2004] found that sexual activity enters strongly in happiness equations and that sex seems to have disproportionately strong effects on the happiness of highly educated people.

I thus propose to use a combination of the three love types, according to their availability, the most potent approach being a combination of the three. Activated as much and as often as possible, both at the sexual, romantic and attachment level, they ensure a flow of hormones and of neuro-transmitters that reward the brain and stimulate it. They contribute to keeping an optimistic and dynamical active attitude. And physical together with emotional love performance is probably one of the best training activities for the brain and the body. It is an excellent catalysis for the body and mind to start a day and/or to end it. It ensures beneficial circulations of nervous stimulations, hormones and oxygenation. Under stress or during tasks that requires stringent attention and concentration, I recommend taking one minute to immerse oneself to empathize with the loved ones, perhaps aided by empathic, family, romantic or sexy pictures or a quick phone call, so as to trigger happiness hormones and well-being. This minute of "relaxation" will boost brain performance for hours. On the long term, lust-love, attraction-love and attachment-love together ensure top performance, longevity with strength and health.

**3-Deep breathing and daily exercises**

Every day, we need to inhale about 35kg of oxygen (and about four times as much air), drink about 2kg of water and eat approximately 1kg of food. The ratios of these weights give a first impression on the relative importance of the three basic inputs our body needs. Another way to look at this is to remark that, while it is possible to survive more than one month just on water, and about a week without water, a few minutes without breathing are fatal. We take breathing for granted, as it occurs via the unconscious control of the autonomous nervous system. But breathing, as we all know, can also be controlled consciously and with great benefits.

Deep breathing and exercises are essential to oxygen the brain, the organ that amounts to 2-3% of the body weight and consumes 30% of its oxygen. Deep breathing and exercises also catalyses the circulation of the lymph system (the body fluid around our cells that carries all kinds of substances, including wastes generated by cells). Deep breathing and exercises also provide an internal massage to all our organs, activates our blood stream and propels the oxygen and nutrients on all parts of the body in a much more dynamical way. In addition to promoting cardiovascular health, preventing diabetes, and lowering cholesterol, compelling evidence has recently shown that routine exercise have a long-lasting beneficial effect on slowing the progression of neurodegenerative diseases, such as Alzheimer's disease and Parkinson's disease (Fryer et al., 2011; Gitler, 2011).

Thus, every morning, with basically no exception, I start my day by approximately 10 minutes of deep breathing and exercises. A 10-minute breathing and exercise session is sufficient to feel energetic and dynamic for the day. When in a hurry, I scale it down to just 3 minutes, but the resulting feeling of energy and well being is worth it. Sometimes, I feel like extending it a bit more to 15-20 minutes but most of the time it is just under 10 minutes, a duration that can be easily inserted in even the busiest schedule (think of this on par with the time that you take for a quick shower and for brushing your teeth). If necessary, cut on a full breakfast and just drink a lot of water and eat fruits (fresh and dry) that you should also get during the morning to perform till lunch. But don't cut on this morning exercise session. In some rare cases, I am late and have to forgo even the 3 minutes. Then, I breath by running to work, by climbing stairs, by contracting my muscles (ankles, legs, ass, abdominals, pectorals, arms, shoulders, and so on) when standing in the tram or while driving. Any idle moment is an opportunity to breath and exercise to fight our sedentary mode of life, which is slowly but surely debilitating us.

It is important to find the best and most pleasant exercises that adapt to your physiology and level of physical fitness. This may evolve. For myself, I stick to the following classics that I want to be able to practice in the bedroom or hotel room when traveling, in the office, anyplace where I can find two square meters free. No need to go to the gym, no lost time, just well being, efficient and performing.

For those with tastes for different styles of exercises, I stress that it is essential to perform some sufficiently intense exercises that make the heart accelerate to about 120-150 pulsations per second and result in the need for deep breathing and some sweating [Ferriss, 2011]. Yoga and other slow static routines contribute an excellent component of the daily training, but are insufficient just by themselves.

I always start with an exercise that combines breathing with stretching of the shoulders and arms. It consists in simultaneous rotations of the two shoulders with arms kept straight, like swimming the backstroke but with the two arms moving in synchrony and remaining straight. The goal is to pull gently on the joints and muscles as much as possible and to open widely the chest, while synchronizing the inspiration with the ascending motion of the arms and the expiration with the descending motion of the arms. I start this first exercise slowly at first, breathing deeply in synchrony, putting my shoulders are far back as possible, and then progressively accelerated both the arm-shoulder motion in synchrony with the breathing. I also open wide my palms stretching their muscles as much as possible during the ascent and relax them during the descent. I typically do 50 to 100 rotations. The brain is then flushed with oxygen. The upper body is inundated with energy provided by the combined deep breathing, opening of the chest, and more strenuous exercises (at the end when the speed is the fastest). This exercise has the benefits of straightening the upper back, put the neck and head upward and providing a powerful pose for the body. Moreover, it ensures that the shoulder muscles, arguably the second most important muscles after the abdominals discussed below, are strengthened. Shoulders exhibit the largest numbers of degrees of freedom in our body. As an engineer, one can imagine the complexity of shoulders. Strong shoulders enable avoiding lots of injuries and changes life, really.

I follow immediately by stretching my legs and back with diverse variants, such as wide opened legs and descending arms and upper body to the floor as much down as possible. Then, same exercises with straight parallel legs with the goal of touching the floor with flat hands or more. Synchronize deep breathing with inhalation during the upward motion and exhalation during the downward motion. This exercise has the triple benefit of (i) working on the stretching of the tendons and muscles on the back of the legs, in particular the harmstrings, (ii) working on the nerves, bones, muscles, ligaments and tendons of the back and in particular the extensor, flexor and oblique muscles in the lower back and (iii) oxygenation.

Then, I work my abdominal muscles (abs), which is essential for the health of my back muscles and vertebra disks. It is well known that back pain is one of the most common types of pain in adults, and by far the most common cause of back pain is back muscle strain. The spine can be compared with the mast of a sailboat, which is held by cables that are positioned both in the front and on both sides towards the back of the boat. The forces that transmit the power of the wind on the sails to the boat hull come from the cables, not from the mast itself. Similarly, the strength of the body results from the interplay between the muscles in the back and in the front, the most important ones being the abdominal muscles. Thus, to have a strong and healthy back, you need to have strong abs. This will prevent one of the curses of the modern sedentary mode of life, which is back pain. The back pain problems experienced at some time in the life of most people will not occur. Even better, you will be able to carry heavy load without thinking about it. As a testimony, I can carry easily 100 kg and more from the floor up, moving heavy furniture or big boulders or tree trunks in my garden. This results simply from the morning session of abdominals that I never miss. Strong abs also ensure a better sitting upward position while working at the desk, which itself slows down fatigue of the back.

I like particularly to perform about 50-100 sit-ups (you can start with 5 or 10 and then progressively scale up), being careful to have the full amplitude motion to work both on the back and on the abdominals. I perform variations with twisting to the right and to the left during the ascent. Again this is done with deep breathing synchronized with the motion. You will see your performance increase remarkable fast from day to day.

I finish with 40-50 push-ups. There are a large variety of styles of push-ups in which you vary the distance between the hands, and their position upward or downward with respect to the center of gravity of the body. For the minimum morning session, I just use the standard version. When I have more time and feel like it, I will use more challenging variants. Ladies can do the push-ups not fully horizontal but pushing from a wall or with hands on the bed or a table with a variable angle that can be adjusted depending on their strength. This gives great arm and breast shapes.

That's it. This takes between 3 and 7 minutes, depending on the variations and intensity.

During the day, to relax my body and oxygenate my brain, I do 50 push-ups for 1 minute and then go back to work. Or I jump in three sessions separated by 30 seconds recuperation: first 20 jumps as high as possible trying to touch the ceiling with my head, then 15 and the last one with another 15. Perhaps, the occupants of the office below mine may wonder about the clap sounds above their heads. This is for the good cause. But, I also try to land softly and supplely. Or, I kick like in karate or kung-fu (of soccer type or as in dancing if you don't know otherwise) with each leg alternatively. This is excellent for the blood circulation in the legs, their stretching and for building their strength. I do about 50 kicks on each leg and several types of kicks. All these exercises are easy for anyone to perform and do not need space or special conditions. Adjust the numbers and intensity to your condition. As you practice more just a few minutes here and there during the day, your level will improve, your well being and performance will follow. This is related to the exercise strategy known as high-intensity interval training, which is intended to improve performance with short training sessions [http://en.wikipedia.org/wiki/High-intensity_interval_training]. I also like to think with Art de Vany [2010], that we should behave and train similarly to our Paleolithic ancestors, jumping from rest positions to escape a predator attack or to chase a prey as a sudden opportunity arises, this being done following "fractal" intermittent patterns [Mandelbrot, 1982].

I practice such short intense workouts at random, when I feel like it. They last just a few minutes, but they essentially reboot my brain. After hours of intense concentration, there are times when you feel tired, or even exhausted. Actually, you are not and it is not rest that you need, because your body has not been really exercising. Your nervous system is tired, your body is clogged and your muscles and tendons are quenched by remaining in the same sedentary positions. You will acquire new energy and rejuvenate your system by these short intense workout sessions that I recommend strongly. It works marvel for me.

I first discovered the benefits of short intense workouts years ago when, as a student, I had to pass a long series of exams spread over one month, with a rather harsh regime of a four hours exam every morning followed by a 4 hours exam every afternoon, five days a week and for four weeks. Under this regime, most students finish exhausted. The first day, I tried meditation and relaxation exercises during the lunch break. The result was disastrous in the afternoon where stress and febrility built up. Indeed, during an exam or any similar intense time of concentration, stress and anxiety as occurs also during competition, the nervous system needs to be relaxed and oxygenated. I claim that the best approach is body exercise. The second day and all the others that followed, I did some of the exercises mentioned above during the lunch breaks. This worked very well for me, as I relaxed completely from the tension of the morning exam and went to the afternoon event essentially as fresh as in the morning. I think this participated significantly in allowing me to be selected in the best "Grande Ecole" of the annual national French competition.

You may want to complement these impromptu super-short sequences by one or two real workouts per week. Personally, I take 40 minutes once or twice a week to train with my son Paul. We start by 20 minutes of slow motions to stretch and bend muscles of all parts of the body. This is done without interruption and intensely so that we already sweat profusely. For the next 20 minutes, we then go to the training room and move weights, by repeating three times without stops two different sets of 4 exercises. Each new session, we change some of the exercises to add diversity and some fun. Another trick is to perform all exercises with complete control and a good position, such as a straight back in all circumstances. A good position ensures that you are working out your muscles optimally and that you are not straining your ligaments, joints and back. You want to conserve your body for decades and exercises have to be done to maximize strength and resilience, to prevent fatigue of organs and body parts as well as aging. It is comforting to see that we are on par with the big guys who spend 2 hours everyday lifting weight but leave the gym almost without needing a shower, a sign of inefficient use of time! This is the secret: intense short workouts.

The general attitude described in this section ensures, I believe, a better preparedness against the risks of accidents and injuries, and especially for the more extreme sports that I practice. The trick is that every effort should be part of a whole virtuous circle. Think of the process, dive in the event with your brain, for instance by using a short check-list including a few flexions, stretching and strength exercises as well as the mental images of what is to come. This way, you become ready for practically anything. In other words, like our hunter-gatherers who lived tens of thousand years ago, I try to keep a kind of "fractal" style of training and mindset [de Vany. 2010]. I claim this minimizes dramatically the risks of injuries, which often occur by surprise for most people.

**4-Water and chewing**

A normal average human body is made of approximately 60% of water. Water is the natural transport fluid involved in all important communication networks of the body. Drinking a lot of water (medium to low mineralization) ensures that the organs responsible for filtering and cleansing have sufficient carrying capacity and can eliminate toxins and wastes. You can see the argument as just plain good sense that we need sufficient water for dilution of wastes. Chemists and physicists know water as a truly exceptional fluid on many accounts. It is not by chance that life emerged and organized itself with water. Water is the best solvent that can dissolve an astonishing number of solid substances. Dehydration is a dominant cause of tiredness, pain and chronic diseases. Drinking about two liters per day plays marvel to keep energy and remove fatigue. Indeed, unknown to or often forgotten by the public is the fact that many chronic diseases may be associated with an insufficient intake of water. Such a simple gesture as drinking water regularly during the day may go a long way towards avoiding fatigue and remaining healthy.

I emphasize drinking pure water and not sodas, cokes, processed juices, coffees and so on. Just plain water. The drinks that are processed by agro-business are in general provided with added sweeteners that are known to promote malignant cell growth [Servan-Schreiber, 2009]. In addition, drinking with sugar (glucose, saccharine) prevents the body from strengthening its metabolic pathway of storing fat in time of surplus and burning fat efficiently in case of need. By feeding our body continuously with sweet drinks and sugar, we saturate our blood with sugar and we weaken considerably the metabolic processes of storing and retrieving sugar, making us more vulnerable to hypoglycemia in the rare cases where external sugar intakes stop. The consequence is to easily feel weak and tired. In contrast, letting the body be just flushed by plain water for hours ensures the build up of its metabolic capacities to burn fat efficiently. This is like muscle build-up by training. Art de Vany [2010] develops convincing supportive arguments for this. He correctly argues that our body is basically inherited from our hunter-gatherer evolution and we are thus adapted to strive in a patchy and varying environment for which our metabolism has derived efficient solutions to the energy flow problem. This backfires with our modern caloric and sugar rich, but nutritionally depleted, foods that are available at little expenditure of energy [de Vany, 2010], in the form of chronic diseases, an on-going so-called epidemic of obesity and many other modern so-called developed country diseases [Campbell and Campbell, 2006].

For the determination of the amount of needed water, a rule of thumb is to keep urine transparent. Do you feel a bit tired? Drink water. The effect is almost instantaneous. I constantly keep one or two liters on my office desk and drink when I feel like it and outside meals. I always carry water with me on trips. A minor nuisance is to drink it all before going through airport security.

Of upmost importance is to drink our water outside meals. Most people use breakfast, lunch and dinner times to fill their body with both the liquid and solid nutrients that their bodies require. This is logical since meals are the times when we re-fuel our body. However, this behavior constitutes a fundamental mistake. Ponder this question: what other mammals in the whole animal kingdom drink their water together with their solid meals? None! We are the only one among about 5500 known mammal species who do it. The convenience of tools and the development of technology have put bottles on our meal tables to consume at the same time we ingest solid food. This apparent gain of civilization collides against a healthy diet for at least three reasons:
(i) Drinking lubricates and help swallow insufficiently chewed morsels; but digestion in the stomach requires the comminution of our food into particles that should be as minute as possible in order to maximize surface over volume ratio and therefore facilitate the digestive chemistry performed by the gastric secretions. This is just plain and simple good sense chemistry. When digesting unbroken food morsels, the stomach and the whole digestive system has to secrete more, takes more time to process our food, all this cumulatively increasing tiredness and fatigue on the body over the long run. I therefore recommend chewing so that you "drink your food". Similarly, water and liquids should stay a while in the mouth before swallowing to warm up and mix with saliva so that you "eat your drinks". A difficult digestion starting in the mouth is probably significant contributor to the feeling of tiredness after a meal.
(ii) Starch and other vegetable substances start their digestion with the help of enzymes found in the saliva; lengthy chewing ensures optimal chemical reactions with these enzymes and saves energy for the rest of the process in the stomach and intestines.
(iii) Ingested fluids dilute the stomach secretions, thus hindering the digestion process. Again, plain and simple good sense chemistry.

**5- Fruits, unrefined products, food combination, vitamin D and sun exposure and no meat and dairy**

Our body is made of the chemical building blocks that we eat. Using these raw materials, our body possesses an extraordinary ability to control finely tuned mechanisms to build, function and repair. But if repeatedly stressed and continuously abused, it is a well-established phenomenon in biology that the body will lose progressively its ability to control its feedback mechanisms. I find the evidence that this disequilibrium may result in chronic disease, cardiovascular diseases, and diverse forms of cancer very compelling [Campbell and Campbell, 2006; Servan-Schreiber, 2009]. Hippocrates, the Greek founder of western medicine, said: "Let your food be your medicine, and your medicine be your food."

Our hunter-gatherer ancestors ate a lot of fresh fruits and plants. All the great apes, with the exception of the gorilla, who is also folivorous (leaf/vegetative-eaters), are primarily frugivorous, while they complement their diets by being opportunic omnivorous with a significant intake of insects and eggs. We should eat a lot of fruits and vegetables in particular because our body has evolved to profit from them [de Vany, 2010]. Fruits are a great way to ingest water as many of them are made up of 80% water in weight. They are full of vitamins, these organic chemical compound that cannot be synthesized in sufficient quantities by our organism and must be obtained from the diet. Many fruits contain natural fibers that help regulate bowel movements. Natural phytochemical antioxidants are abundant in fresh vegetables and fruits. Recall that antioxydants remove free radicals, which are believed to be involved in degenerative diseases, senescence and cancers. The fruits that provide the largest number of antioxidant types are those with dark skin color, with the purple-blue-red-orange spectrum being home to the most antioxidant-rich fruits. Many medical studies have demonstrated the benefits of eating plenty of fruits and vegetables, with provide the benefits of preventing heart disease and stroke, of controlling blood pressure, of avoiding some types of cancer, and of fending off cataract and macular degeneration, among many others.

A great way to eat fruit is to drink freshly squeezed fruit juice. Personally, my first drink of the day consists of 2-4 oranges worth of freshly squeezed juice. I have also found that oranges are much better digested in the morning before any other intake. To avoid conflicts with orange acidity, I usually wait about 15 minutes before eating other fruits, which are sweet or sub-sweet such as apples, melons, strawberries, berries, kiwis and so on. Then, 15 minutes later, if I am still hungry, I eat nuts (almonds, figs, raisons, and so on), or cereals or whole wheat bread with honey. I do this in my office. If time is short, I just drink the freshly squeezed orange juice or, if a juicer is not available during my travels, I eat fruits. My body then feels strong, energetic and light at the same time.

The rule that acidic and sweet fruits should not be mixed extends to all foods. Indeed, one should avoid mixing foods requiring an acidic medium with foods needing alkaline juices in order to be digested. One should also avoid mixing foods that require widely different speed for digestions. Again, this is just obvious chemistry and physics. Think of our digestive system as a complex chemical plant. Let me dare the following analogy: the way we mix foods in modern meals is akin to requiring that a chemical oil refinery should be able to process in the same unit both heavy viscous tar oil with light crude oil together with purifying water for drinking and transforming coal into benzine! For an industrial unit, this is clearly impossible. But this is the kind of task that we require our body routinely to perform. Being highly adaptive, our digestive system does much better than any oil refinery and complies with our erratic and irrational or irresponsible unconscious behaviors, but at the cost of tapping our energy and of suboptimal absorption of the vital elements.

The fastest foods are, apart from water followed by water-filled fruits, simple sugars followed by starch (carbohydrate), proteins and fats. Eating something that takes a long time to digest, followed by another food that would normally digest very quickly alone, forces the later to stay with the former to digest, simply because there is no mechanism for demixing the foods that we have thrown all together in our stomach and in our intestine. In such situations, the sugar and starch will ferment while protein putrefies, with inconveniences that range from uncomfortable feelings, tiredness, to serious diseases over the long-term.

Concerning the issues of acidic versus alkaline, one needs to know that starchy foods (wheat, bread, rice, cereals, potatoes) require an alkaline medium for digestion, which is associated with the ptyalin enzyme found in the saliva (hence the importance of chewing starchy foods). In contrast, breakdown of proteins requires an acidic medium. It is an inevitable law of chemistry that digestive medium cannot be both acidic and alkaline. It is therefore highly recommended to avoid incompatible mixing. The classification of food in acidic versus alkaline can be found in standard references and the consultation of their summary tables should suffice to avoid most common mistakes, and get the most energy and metabolic benefits from our foods [Shelton, 1940; Marsden, 2005].

Last but not least, I recommend avoiding animal products, especially meat and dairy. The "China Study" [Campbell and Campbell, 2006] has compiled an enormous volume of evidence showing that ingestion of meat, especially read meat, as well as cow milk is strongly detrimental to health. Controlled experiments on rats have shown that tumors, induced by ingestion of carcinogenic substances like dioxin, are striving in meat-fed animals while they recede in vegetable-fed animals. The "China Study" compares the long-term health characteristics of 65 chinese groups with those of western countries and demonstrates that many disorders are rich-world diseases in large part created by bad feeding diets. Chineses, who were immune to the chronic diseases and other afflictions of western societies and who emigrate to the West or simply adapt the western habits, develop progressively their diseases, demonstrating that the difference is not essentially genetic but mainly environmental and nutrition based. The "China Study" shows that cognitive disorders (dementia), kidney stones, heart disease, obesity, multiple sclerosis and other autoimmune diseases such as rheumatoid arthritis and type 1 diabetes are all markedly reduced by plant-based diet. Animal protein, even more than saturated fat and dietary cholesterol, raises blood cholesterol levels in experimental animals, in individual humans and entire populations [Campbell and Campbell, 2006].

A boost of well-being is obtained by just being bathed by sunlight. Whenever I can, I enjoy exposing some part of my skin to sunlight, not for hours as some beach lovers do, but for about 15 minutes or less. This is sufficient for our body to synthetize enough vitamin D. The vitamin D is stored in the liver in fat with a residence time of several weeks. Thus, intermittent sun exposure is fine as our body can store vitamin D. When needed for anti-oxydation and for the functioning of the immune system, it is transformed into a supercharged form called 1,25 D, which lasts only a few hours. It turns out that animal protein intake blocks the formation of 1,25 D in the kidney. This limits the amount available to maintain and keep the normal cells from deteriorating into diseased cells, and to regulate calcium levels [Campbell and Campbell, 2006].

In summary, an enormous body of scientific literature shows and thousands of different studies demonstrate a protective benefit of plant-based foods and/or harmful effects of animal-based foods for many different diseases. A plant-based diet therefore constitutes in my opinion the foundation of performance.

**6-Power foods**

In addition to the simple above recommendations, let me share some of the "tricks" I use to face stressful situations.

When I have to travel around the world, I have sometimes to spend 10 hours per day of intense concentration needed to interact with very intelligent people, to speak to them, teach and present vigorous messages. Sometimes, the challenge is to talk as intelligently as possible and keeping inspired and creative non-stop for +10 hours per day for a full week, while traveling around the world with jetlag. To face the physical and cognitive demands, I have found that emphasizing a combination of fresh and dry fruits is especially efficient. Before a high-demand event (and this can be several times a day), I typically ingest dry figs, apricots and dry almonds, as much as my body requires. This works marvelously for me. In addition, some pieces of dark chocolate (>70% cacao) are also great for the brain (magnesium) and for the mood (in addition to the taste, chocolate contains serotonin that acts as an anti-depressant). Dark chocolate health benefits stem in part from flavonoids, the antioxydants that are especially high in cocoa and which neutralize oxygen radicals.

I am not entirely immune to flu or colds, especially after I have pushed my body a bit too much, the tiredness cascading to a weakened immune system. When I feel a slight decrease of my vitality and/or a tendency to sneezing or the beginning of a sore throat, I immediately lighten my diet, drink even more water and, here is the most important, I start eating raw onions and garlic. Onion is a natural antibiotic rich in sulfur and flavonoids. It is known to be one of the best natural remedies to combat infectious diseases of the respiratory tract. Chewing raw onion so that the respiratory track, the throat and the sinuses are impregnated with its flavor brings me almost instantaneous relief. Since I started practicing this trick, I have been able to recover my full performance in no more than 24 hours. Onions are also great in general in our diet because they are full of vitamins, such as C, B1 and B6, along with potassium, phosphorus and fiber. George Washington, the founding father of the USA, has been quoted as follows: "My own remedy is always to eat, just before I step in to bed, a hot roasted onion, if I have a cold." My own experience is that raw onion is even more active and efficient. However, you have to convince your significant other and family (pets are great because they do not mind!) to do the same, so that the inconvenient smells cancel out and the benefits of love, cuddling and sex can still be garnered!

Similarly for garlic, which can help cure bronchitis and sinus infection. Since Louis Pasteur observed garlic's antibacterial activity in 1858, garlic has been found to have antibacterial, antiviral, and antifungal activity in vitro and to prevent heart disease (including atherosclerosis, high cholesterol, and high blood pressure) and cancer. My own experience is that raw garlic fluidifies my blood and provides a great feeling of well being. Lemon (in particular for its ascorbic acid, also called vitamin C), dry nuts and dry fruits, and kiwis are among other foods that my body enjoys particularly to feel vigorous.

My last advice is to listen to your body, to the weak but omnipresent signals that your stomach, intestine, heart and overall nervous system emit. For instance, I have found that I can tell if a food is good for me by listening to how my stomach and overall body reacts to the first mouthful. Our body and mind constitute complex entangled networks of sensors feeding the brain, both conscious and unconscious parts, with many channels of information for continuous feedbacks. The boundary between conscious and unconscious is porous and I believe that training and meditation can enhance our performance by augmenting our internal sensory perceptions. Hippocrates said: "If you are not your own doctor, you are a fool."

**7-Play (homo ludens), intrinsic motivation, positive psychology and will**

Many animals play but perhaps the most distinguishing of our characteristics is that we are gifted to play. Some years ago, I coined the term "thirst-for-play" to embody the hypothesis that we have evolved as cooperative social animals with an extraordinary drive to play (but many of us forget it in their adulthood, forced by the serious facts of life). Look at kids, whose goal in life seem to be playing all the time all sorts of games (and more and more video games in the electronic era). Play is motivated by (a) the desire for novelty, for changes, for new experiences and (b) the drive to "flex one's muscle", skill and abilities and push them to the fullest of their extent. When playing favors cooperation, then cooperation strives. Team sports are good example. But take the open-source phenomenon: programmers spent an apparently unreasonable fraction of their time working on difficult problems without being paid. One explanation advanced based on surveys is that programmers enjoy it (it is like an adult video-game) as they can create (like a lego) and they take pleasure in their quest for adventure (experience of exploring new software territories) that is useful for them, and they can flex their programming muscle (in addition to reputation building). As adults, we tend to lose our ability to play as we did as children. I personally feel very lucky to be able to see my profession as an opportunity to "play" essentially 100% of my time. This does not mean that I take it superficially, on the contrary, as being in a playing mood enhances the acuity of attention and the drive for performance. By embracing each so-called work activity as a "play" makes life worthwhile. In addition to playing, some relaxing with a blanked out mind, taking the bike for a small ride, go for a walk because you feel like it, chit-chat with friends, family, kids, make fun of people, joking, are all excellent to de-stress and rejuvenate energy.

High-level performance is first a child of the mind and foremost a result of high intrinsic motivation. Intrinsic motivation is about finding for oneself a goal, something worth living for and/or working for, passionately. For the genuine real world movers, it is much more than monetary incentives. Many pieces of evidence [Fleming, 2011] together with meta-analyses [Deci et al., 1999] question the validity of the standard axiom of economists and workplace human resources that people are essentially motivated by extrinsic rewards. In fact, in virtually all circumstances in which people are doing things in order to get rewards, it has been found that "extrinsic tangible rewards undermine intrinsic motivation." This often leads to unwanted long-term value destruction illustrated by the creative accounting of the likes of Worldcom and Enron, as well as by the unethical behavior and dismal performance of "Wall Street" individuals and firms again revealed since the unfolding of the 2008 financial crisis.

I believe that intrinsic motivation is like love. It can come like a striking bolt "at first sight" or nucleate first timidly, then grow progressively through a long careful nurturing process to fully blossom. Intrinsic motivation is the result of processes with positive feedbacks, the more one practices internal motivation, the more it grows. A bit like happiness, which according to Abraham Lincoln is such that "People are just as happy as they make up their minds to be," intrinsic motivation is a state of mind that grows with practice.

I often hear around me "I do not like to…" or "If I do this or that, I am afraid to fail", or "I do not know how to…" Most people would not even express these phrases aloud, but these weight internally in their minds. My reaction is to replace these negative expressions that reveal a losing attitude by "I like to try it because I can learn new things," or "When doing this or that, I will be focused and brave," or "This is an opportunity for me to start learning how to…" By forcing a positive constructive attitude, behaviors can change and improve. I believe that it is possible to reprogram some of our software acquired during our formative years and transcend fears and limitations. This touches upon the limits of present scientific knowledge on the brain as an organ that supports the cognitive and psychological processes that define us. There are no universally accepted techniques for improving and programming the brain for higher performance. But, by thinking of our mental processes are resulting from software-like constructions acquired during our lives, I believe it is possible to have an influence and modify the unwanted code bits, similarly to an computer engineer debugging and improving a computer program. One has to listen again to the weak signal and the feedbacks offered by our brain, and to experiment individually. Meditation is a helpful way to "reset", by listening to the feedbacks that our body and mind send continuously. This empowers us, provides a sense of relevance and importance.

The faculty of Will is a key driver of performance. How to develop your will? I believe that, like strength and health, will is a quality that can be nurtured and developed by a slow cumulative process of small gains that grow progressively via the action of virtuous positive feedbacks. Will can be catalyzed by an optimistic positive psychology, which can profit from the existence of a social network with high emotional quality to provide the essential needed feedbacks. Intrinsic motivation can blossom when freedom to think and to act provides a sense of control on one's life and lead to the recognition that what happens to us is our responsibility. Success smiles to the one prepared to react opportunistically to chance.

**As a conclusion:** The great people, the geniuses and the innovators, who have changed something in our World, are characterized by remarkable levels of performance. But all too often, I have observed a disconnection as they did not care or did not realize that the first machine/creature/entity to care about is their body. In order to achieve something or to help others, one needs to be strong and to be an example. I have seen too often these great people pushing their biological machine to unsustainable levels, over-loading it by stress and poisons. Too many great people have left us much too early, succumbing to debilitating diseases, cardiac diseases, cancers and the like, as a result of their falling prey to this blind spot. I hope I have excited your curiosity about the possibility to achieve one's full performance potential for the long term, by cultivating a personal hygiene that is a reflection of one's goals.

**Acknowledgements:** My son Paul-Emmanuel and my brother Thierry have provided enthusiastic support and feedbacks on several parts, in particular on exercises/training and feeding. I am also grateful to my research collaborators Susanne von der Becke and Peter Cauwels at ETH Zurich for useful feedbacks on a preliminary version.

**References**

Ader, R., D. Felten and N. Cohen (1990) Interactions between the brain and the immune system, Annu. Rev. Pharmacol. Toxicol. 30, 561-602.

Blanchflower, D.G. and A.J. Osward (2004) Money, sex and happiness: an empirical study, Scandinavian Journal of Economics 106 (3), 393-415.

Campbell, T.C. and T.M. Campbell (2006) The China Study (the most comprehensive study of nutrition ever conducted and the startling implications for diet, weight loss and long-term health), Bendella Books, Dallas, Texas.

de Vany, A. (2010) The evolutionary diet (What Our Paleolithic Ancestors Can Teach Us about Weight Loss, Fitness, and Aging), Rodale Books.

Deci, E.L., R. Koestner and R.M. Ryan (1999) A meta-analytic review of experiments examining the effects of extrinsic rewards on intrinsic motivation, Psychological Bulletin 125 (6), 627-668.

Ferriss, T. (2011) The 4-hour body (an uncommon guide to rapid fat-loss, incredible sex and becoming superhuman), Vermilion (Imprint of Ebury Publishing, part of Random House Corp), New York.

Fleming, N. (2011) The bonus myth: How paying for results backfires, New Scientist 2807, 40-43 (9th April).

Fisher, H. (2004) Why We Love: The Nature and Chemistry of Romantic Love, Henry Holt.

Fisher, H. (2009) Why Him? Why Her? Finding Real Love By Understanding Your Personality Type, Henry Holt USA-Canada.

Fryer, J.D., P. Yu, H. Kang, C. Mandel-Brehm, A. N. Carter, J. Crespo-Barreto, Y. Gao, A. Flora, C. Shaw, H.T. Orr and H.Y. Zoghbi, Exercise and Genetic Rescue of SCA1 via the Transcriptional Repressor Capicua, Science 334, 690-693 (2011).

Gitler, A.D., Another Reason to Exercise, Science 334, 606-607 (2011).

Mandelbrot, B.B. (1983) The Fractal Geometry of Nature, W.H. Freeman, San Francisco.

Marsden, K. (2005) The Complete Book of Food Combining: A New, Easy-to-Use Guide to the Most Successful Diet Ever, Piatkus Books.

Osorio, I., M.G. Frei, D. Sornette and J. Milton (2009) Pharmaco-resistant seizures: self-triggering capacity, scale-free properties and predictability? European Journal of Neuroscience 30, 1554-1558.

Osorio, I., M.G. Frei, D. Sornette, J. Milton and Y.-C. Lai (2010) Epileptic seizures, quakes of the brain? Physical Review E 82 (2), 021919.

Robbins, T. (1997) Unlimited Power: The New Science Of Personal Achievement, Free Press.

Savage, V.M. and G. B. West (2007) A quantitative, theoretical framework for understanding mammalian sleep, Proceedings of the National Academy of Sciences (USA) 104 (3), 1051-1056.

Servan-Schreiber, D. (2009) Anticancer: A New Way of Life, Michael Joseph.

Shelton, H.M. (1940) Food combining made easy, Willow Pub; 1st rev. print edition.

Sornette, D., F. Ferré and E.Papiernik (1994) Mathematical model of human gestation and parturition: implications for early diagnostic of prematurity and post-maturity, Int. J. Bifurcation and Chaos 4 (3), 693-699.

Sornette, D., V.I. Yukalov, E.P. Yukalova, J.-Y. Henry, D. Schwab, and J.P. Cobb (2009) Endogenous versus Exogenous Origins of Diseases, Journal of Biological Systems 17 (2), 225-267 (http://arxiv.org/abs/0710.3859)

Wagner, U., S. Gais, H. Haider, R. Verleger and J. Born (2004) Sleep inspires insight, Nature 427, 352-355.